\documentclass[aps,prl,preprint,superscriptaddress]{revtex4}
\usepackage{rotating}
\usepackage{amsmath}
\usepackage{graphicx}
 \usepackage{amssymb}
 \begin{document}       
 
\title{F. J. Dyson: The Man  who would make Patterns and Disturb the Universe}
\author{Patrick Das Gupta}
\affiliation{Department of Physics and Astrophysics, University of Delhi, Delhi - 110 007 (India)}
\email{pdasgupta@physics.du.ac.in}
\maketitle

\section*{Abstract}
Freeman J. Dyson, a brilliant theoretical physicist and a gifted  mathematician, passed away on 28 February 2020 at the age of 96.  A vignette of his outstanding contributions to physical sciences, ranging from the subject of quantum electrodynamics to gravitational waves, is provided in this article. Dyson's futuristic ideas concerning  the free will of `intelligent life' influencing the remote future of the cosmos with `Eternal Intelligence', Dyson tree, Dyson sphere and so on, have also been discussed briefly.
\section*{}
(This article has appeared in Resonance, Vol.25, No.10, pp.1319-1337 (2020))
\section {Introduction}

Let me begin with some curious observations:  A brilliant young person  after publishing  research papers of significance in pure mathematics, moved over to  physics and, within a year or so,  published  a far reaching article that would continue to guide future theories. This gem of a research paper  synthesized, in a very elegant manner,  discordant formalisms  propounded  independently by three other outstanding theoretical physicists, who later shared the Nobel Prize in physics for their pioneering work.

With time,  our  bright and innovative  protagonist kept trotting  peripatetically in the realm of physical sciences from one field to another. Several  distinct scientific ideas are named after him in the area of physical sciences.  To top it all, this individual never had an official Ph.D. degree.  How often, in the history of science,  does one encounter such a sequence of events? 

The immensely gifted  person, I am referring to, is none other than -   Freeman John  Dyson, who sadly left this world  on February 28, 2020. Freeman Dyson was born in England to Mildred Lucy Atkey and George Dyson on December 15,  1923. It is generally acknowledged that Dyson's father, Sir George,  had been   a  very talented composer  as well as a teacher-musician. 

Freeman Dyson excelled in his school studies and, at the age of  twelve,  he moved  to Winchester College after winning the first position in  a scholarship test in 1936. While he was vigorously pursuing mathematics, concentrating on his favourite subject, number theory, Dyson had already studied Eddington's `The Mathematical Theory of Relativity' by  1939.

In 1941, Dyson was offered a scholarship to study in the Trinity College, Cambridge, where he was taught by  famous mathematicians like the legendary  G. H. Hardy. In Cambridge, he also studied physics under one of the greatest theoretical physicists, P. A. M. Dirac. After the World War II, his mentors at Cambridge advised him to take up physics. In particular, the renowned fluid dynamics expert, Sir    
Geoffrey Ingram Taylor,  penned  a letter of reference to Hans Bethe, a brilliant theoretical physicist and later a Nobel Laureate, of Cornell University, USA, containing  the following lines:

"You'll have received an application from Mr. Freeman Dyson to come to work with you as a graduate student. I hope that you will accept him. Although he is only 23 he is in my view the best mathematician in England."

Dyson joined Hans Bethe in 1947, and thus began his adventures in physics. Bethe, during that time, 
 inspired by H. Kramer's idea of `renormalization' that the measured  energy of a charge particle is a sum of its bare energy and the self energy acquired by its interactions with the electromagnetic field generated by itself,  had published a non-relativistic calculation to explain  tiny shifts of the energy levels of a hydrogen atom that was seen in the Lamb-Retherford experiment but was not predicted by Dirac's relativistic electron-positron theory [1-3]. 

Dyson lost no time in using his mathematical talents to perform a relativistic calculation (while ignoring the intrinsic spin of electrons) to show that the shifts match accurately with experimental results. This work was received  by the editors of Physical Review on December 8, 1947, and was published in 1948 [4]. Incidentally, around the same time, Julian Schwinger too had published a relativistic calculation to explain the  Lamb-Retherford shift  of energy levels [5].

 Bethe  by then had  recognized how  prodigiously gifted Dyson was, and he went all the way to  convince J. Robert Oppenheimer of the Institute for Advanced Study, Princeton, that  Dyson be accepted in the Institute. Meanwhile, Dyson was discussing with Feynman as well as Schwinger to learn about their distinct formulations concerning interaction of radiation with charge particles.  In 1948, Dyson joined the Institute for Advanced Study, where stalwarts like Einstein, John von Neumann and Kurt G$\ddot{\mbox{o}}$del were deeply immersed in researching on the fundamental aspects of physical and mathematical sciences. 

When  Feynman shifted  to California Institute of Technology in 1950,  Dyson was offered Feynman's professorship at the Cornell University, Ithaca.  Dyson, after  spending about three years in Ithaca,  returned to  the Institute for Advanced Study  as a professor in 1953, and continued his research studies  at Princeton, writing prolifically on various topics, till he breathed his last on February 28, 2020. 

\section{Dyson and QED}

Quantum electrodynamics (QED) is a relativistic quantum  theory of electromagnetic field, charge particles and their   interactions. QED describes, with great precision, myriads of physical processes involving electrons, positrons and   photons - creation and annihilation of electron-positron pairs, high energy scattering between electrons, positrons and photons, vacuum polarization that causes the observed Lamb-Retherford shift in hydrogen atoms, etc. Vacuum polarization can be understood intuitively as screening of the electric charge of the atomic nucleus due to  spontaneous creation of virtual electron-positron pairs that is taking place continuously in accordance with the uncertainty principle $\Delta E. \Delta t \gtrsim  h $.

 Dyson's most extraordinary contribution to physics  is a paper  that bears the title `The Radiation Theories of Tomonaga, Schwinger and Feynman' in which, in one bold stroke,  he unified the operator-centric Tomonaga-Schwinger formalism with  Feynman's intuitively more appealing space-time approach to QED that utilized propagators in relativistic quantum mechanics [6].   Dyson's article was  received by the editors of the Physical Review journal on October 8, 1948, and was published in the year 1949 on February 1. The basic method to synthesize  diverse viewpoints on QED came to the Greyhound bus passenger, a sleepy Dyson,   in a flash, as narrated by him in a letter to his parents [7]:

September 14, 1948:

"...
On the third day of the journey a remarkable thing happened; going into a sort of semi-stupor as one does after forty-eight hours of bus riding, I began to think very hard about physics, and particularly about the rival radiation theories of Schwinger and Feynman. Gradually my thoughts grew more coherent, and before I knew where I was, I had solved the problem that had been in the back of my mind all this year, which was to prove the equivalence of the two theories.
..."

The sheer elegance, clarity and sincerity of Dyson's paper stare at one's face. The sequence of  names of the physicists appearing in the title is significant. The research paper of Sin-Itiro Tomonaga, indeed was the first one on the subject, followed by the published works  of Tomonaga and his co-workers, of Julian Schwinger and then of Richard Feynman  [8-17]. Dyson endeavoured to emphasize, in a footnote of his celebrated paper, that Tomonaga and his collaborators had an unequivocal head start in building QED [6]. 

For a proper understanding of QED, it is of course necessary  to study  the standard  literature on the subject that is  readily available. Dyson's 1949 paper,  which is remarkably  pedagogical and crystal clear in its approach, may form an excellent supplement [6]. To appreciate Dyson's paper as well as to motivate young readers to study it,  a  fleeting glimpse of the basic pre-requisites has been provided below.

It is well known that  fundamental particles not only have intrinsic spin angular momentum  they must also respect special theory of relativity. In the framework of relativistic quantum mechanics, spin 1/2 free fermions with rest mass $m$ are described by the Dirac equation,
\begin{equation}
\bigg [i\hbar \gamma^\mu \frac {\partial}{\partial x^\mu} - m c\bigg ] \Psi (x^\alpha) = 0
\end{equation}
where the index $\mu$ is being summed over $0,1,2,3$  in the product involving the partial derivatives and the Dirac matrices,  $\gamma^i \equiv \beta \alpha^i$ for $ i=1,2,3$ and $\gamma^0 \equiv \beta $. (The Einstein summation convention wherein repeating indices are assumed to be summed over,  has been used throughout this article.) 

In  eq.(1), $\Psi (x^\alpha)$ represents a column vector with four complex functions $\psi_k  (x^\alpha),\ k=1,2,...,4$ as entries, describing the spin as well as the anti-particle degrees of freedom,
\begin{equation}
\Psi (x^\alpha) =
\begin{pmatrix}
 \psi_1  (x^\alpha)\\
 \psi_2  (x^\alpha)\\
 \psi_3  (x^\alpha) \\
 \psi_4  (x^\alpha)
 
 \end{pmatrix}
 \ \ \ \ .
\end{equation}
Dirac equation can be used to study electrons moving with relativistic energies. However, eq.(1) admits both positive energy $E_{+} \geq m c^2 $ and negative energy $E_{-} \leq - m c^2 $ solutions. Presence of negative energy solutions led Dirac to predict existence of positrons. In the case of plane wave solutions of eq.(1), one may pull out the time-dependent parts  and express them as,
\begin{equation}
\Psi_{+} (x^\alpha)= e^{-\frac{i}{\hbar}E_{+} t} \  \mathcal{U}_{+  \vec{p}} (\vec{r})\ \ \ \mbox {and}\ \ \   \Psi_{-} (x^\alpha)= e^{-\frac{i}{\hbar}E_{-} t} \  \mathcal{V}_{-  \vec{p}} (\vec{r})
\end{equation}
Using the superposition principle, since eq.(1) is linear, general  wavepackets for a free spin 1/2 particle can be constructed out of the plane wave solutions of eq.(3). However, relativistic quantum mechanics limiting to just a single spin 1/2 particle is fraught with problems, e.g.  Klein paradox.

One can show that combining special relativity with quantum theory necessitates a quantum `many particle' description . As an example in 1-dimension [18], if one localizes a particle of rest mass $m$ to a region of Compton wavelength size $\Delta x \sim h/mc$, the uncertainty in the momentum $\Delta p \gtrsim  h/\Delta x = m c$. But then, the uncertainty in energy is $\Delta E \sim c \ \Delta p > m c^2$. Hence, the energy uncertainty is large enough to create extra particles. 

A natural framework to describe relativistic, `many particle' quantum systems is the theory of quantum fields. For instance, in a quantum field theory (QFT), $\Psi (x^\alpha)$ appearing in the Dirac equation is  elevated to an operator status, $\hat{\Psi} (x^\alpha)$, and the Hamiltonian $\hat{H}_\Psi$, which is the energy operator, is constructed out of the Lagrangian that leads to eq.(1). The particle states corresponding to electrons and positrons are simply the eigenstates of  $\hat{H}_\Psi$.

In the classical electromagnetic sector, the time evolution of free electric field $E^i =  F^{0i}$ and magnetic field $B_i =  \epsilon_{ijk} F^{jk}$ are obtained by solving the Maxwell equations,
\begin{equation}
\frac {\partial F^{\mu \nu}} {\partial x^\nu} = 0 \ ,
\end{equation}
 $F_{\mu \nu}(x^\alpha)$ being the electromagnetic field tensor given by,
\begin{equation}
F_{\mu \nu} = \frac {\partial A_\nu} {\partial x^\mu} - \frac {\partial A_\mu} {\partial x^\nu}
\end{equation}
 where $A_\mu (x^\alpha)$ is the electromagnetic (EM) potential with  time-component $A_0$  being the usual scalar potential $ \phi (\vec{r},t)$ while the space components $A^i,\ i=1,2,3$ constitute the vector potential $\vec{A} (\vec{r},t)$.
 
In the quantum domain,  $A_\mu (x^\alpha)$ is turned into a field operator $\hat{A}_\mu (x^\alpha)$. Naturally, the electric and magnetic fields too become space-time dependent operators, and it is their eigenvalues that can be measured at different points  as the observable field strengths. Spin 1 photon states emerge out of the QFT of EM radiation as the energy eigenstates of the Hamiltonian $\frac{1}{8\pi}\int{(\hat{E}^2 + \hat{B}^2) d^3r}$.
 
So far we have described only free fields. But one must include interaction of charge particles  with    EM fields in the theory. The mathematical form of the coupling between a charge particle and EM  fields emerges very naturally when one demands the full theory to be  gauge invariant. In case of the EM fields, it is well known that 
given an arbitrary but smooth  function $\chi (\vec{r},t)$, the electric and magnetic fields remain the same  under the following gauge transformations, 
\begin{equation}
\vec{A}(\vec{r},t) \rightarrow \vec{A}^\prime (\vec{r},t)=\vec{A}(\vec{r},t) + \nabla \chi
\end{equation}
and,
\begin{equation}
\phi (\vec{r},t) \rightarrow \phi^\prime (\vec{r},t) = \phi (\vec{r},t) - \frac {1}{c} \frac {\partial \chi}{\partial t} \ ,
\end{equation}
so that the Maxwell's equations are covariant not only under Lorentz transformations but they are also under the above gauge transformations. 

Going back to the relativistic quantum mechanics described by eq.(1), if one demands that the Dirac equation be covariant under,
\begin{equation}
\Psi (\vec{r},t) \ \rightarrow \ \Psi^\prime (\vec{r},t) = e^{i q\frac { \chi(\vec{r},t)} {\hbar c} } \Psi (\vec{r},t)\ \ ,
\end{equation},
as well as under eqs.(6) and (7), it is necessary that the Dirac equation   be of the form,
\begin{equation}
\bigg [i\hbar \gamma^\mu \bigg (\frac {\partial}{\partial x^\mu} + \frac {i q} {\hbar c}\  A_\mu (\vec{r},t)\bigg ) - m c\bigg ] \Psi (x^\alpha) = 0
\end{equation}
(For a simple exposition to the details related to gauge invariance in the context of Schr$\ddot{\mbox{o}}$dinger equation, one may read [19].)

The above equation  describes  a spin 1/2 fermion carrying an electric charge $q$  and interacting with the EM potential $A_\mu$.  The wavefuntion $\Psi(\vec{r},t)$ at any time $t$  can also be obtained using a technique very similar to the Green's function method,  given any initial state $\psi(\vec{r},t_0)$, 
\begin{equation}
\psi(\vec{r},t) = \int{ K(\vec{r}, \vec{r^\prime}; t, t_0) \psi(\vec{r^\prime},t_0) d^3 r^\prime}
\end{equation}
if the propagator $K(\vec{r}, \vec{r^\prime}; t, t_0)$  corresponding to  eq.(9) has already been determined. In general, one  determines the free propagator (i.e. when $A_\mu=0$ everywhere in the space-time manifold) and then use perturbation theory  to take into account interactions. 

 The propagator method was championed by Feynman in which various interaction processes like vacuum polarization, pair creation, Bhabha scattering, etc. were calculated quite  successfully [15-17]. Feynman  used a space-time approach while imposing consistency with the tenets of quantum theory which ordains that one must include  superposition of all probability amplitudes corresponding to the intermediate processes that  cannot  be observed. Feynman's paper on the theory of positron, submitted after Dyson's 1949 publication,  of course acknowledges Hans Bethe and Dyson for the fruitful discussions [16].
 
 To get a gist of  the space-time approach, one may consider a very simplified description of  an electron getting scattered by a photon. Feynman's method would use the probability amplitude $\psi_{AB}$  for an electron to freely propagate from point A to B by employing the propagator of the free Dirac equation (eq.(1)), and then consider a contact interaction with  the EM potential $A_\mu  $ representing an incoming photon state at point B using a suitable expression determined from the interaction term in eq.(9) and take latter's product with $\psi_{AB}$ to obtain a virtual state that goes from B to C.
 
  Finally, the probability amplitude  for the electron as well as an outgoing photon to  freely travel from C is calculated by taking the inner product  with the amplitude corresponding to the virtual state. Since,  points B and C where  the interactions have been considered are arbitrary and cannot be observed in experiments, the net probability amplitude for the electron-photon scattering is arrived at by integrating over all possible B and C  as demanded by the quantum principle of superposition.
 
 Subsequently, following St$\ddot{\mbox{u}}$ckelberg's interpretation that a  positron can be thought of as a electron going backwards in time [20], Feynman could calculate scattering cross-section of electron-positron interactions  (i.e. Bhabha scattering) correctly [16]. There is a simple way to see why in Dirac's hole theory, a positron may also be interpreted as an electron for which time is running backwards. In the standard viewpoint,   $ t E_{-} < 0$ for negative energy solutions since $  E_{-} < 0$ and $t > 0$ (eq.(3)).  But one may also interpret $ t E_{-} < 0$  as due to energy being positive (as in the case of electrons) but time $t$ instead is negative since its direction is reversed. The space-time approach has been more enduring and appealing as it enabled one to quickly visualize the terms that need to be calculated for a given physical process.  
 
However, unlike the case by case technique developed by  Feynman, the formalisms developed independently by Tomonaga and Schwinger,  are  mathematically  more rigorous, elegant and complete. They are essentially  relativistically covariant formulations involving  field operators corresponding to spin 1/2 fermions interacting with EM fields [8-14]. To get a flavour of the operator based field theory aspects, one needs to start from the Lagrangian density approach to classical field theory as illustrated  below for the spin 1/2 fermions. 

Since $\Psi $ is a column vector field, using the adjoint operation, one may define a row vector field,
\begin{equation}
\bar{\Psi} (x^\alpha) \equiv \Psi^\dagger (x^\alpha)\ \gamma_0 = \Psi^\dagger (x^\alpha)\  \beta
\end{equation} 

Then, eq.(9) can be derived by employing calculus of variation from the following Lagrangian density,
\begin{equation}
\mathcal{L}_D = \bar{\Psi} (x^\alpha) \bigg [i\hbar \ \gamma^\mu \ \bigg (\frac {\partial}{\partial x^\mu} + \frac {i q} {\hbar c}\  A_\mu (\vec{r},t) \bigg ) - m c\bigg ] \Psi (x^\alpha) \ \ .
\end{equation}
One may express the above Lagrangian density as a sum of the free Lagrangian density $\mathcal{L}_{D0} $ that only involves the fermionic degrees of freedom and an interaction term containing the coupling between the charged fermions and the EM potential,
\begin{equation}
\mathcal{L}_D \equiv \mathcal{L}_{D0} + \mathcal{L}_{int}
\end{equation}
where,
\begin{equation}
 \mathcal{L}_{D0} \equiv \bar{\Psi} (x^\alpha) \bigg [i\hbar \ \gamma^\mu \ \frac {\partial}{\partial x^\mu}  - m c\bigg ] \Psi (x^\alpha) \ \ \ \ \mbox{and}\ \ \ \\
\mathcal{L}_{int} \equiv  - \frac{1} {c}\ j^\mu \ A_\mu  \ \ , 
\end{equation}
$j^\mu (x^\alpha) =  q\ \bar{\Psi} (x)\  \gamma^\mu\  \Psi (x) $ being the 4-current density.

Now, in the framework of classical non-relativistic mechanics of point particles, one begins with a Lagrangian $L(q, \dot{q},t)$ corresponding to a particle and obtains the canonical momentum $p \equiv \frac{\partial L}{\partial  \dot{q}}$. Using the particle's canonical momentum $p$, one arrives at the classical Hamiltonian $H(q,p,t)= p  \dot{q} - L $.

In the standard canonical quantization scheme, if one wishes to quantize this system,   one  turns $q$ and $p$ into linear operators  $\hat{q}$  and $\hat{p}$, respectively, and imposes the commutation relation,
\begin{equation}
[\hat{q} , \hat{p}] \equiv \hat{q} \hat{p} - \hat{p} \hat{q} = i\ \hbar
\end{equation}
so that one possesses a quantum Hamiltonian operator $\hat{H}(\hat{q}, \hat{q},t) $ (i.e. the energy observable) starting from the classical Lagrangian $L(q, \dot{q},t)$.
 
An analogous program is carried out in the case of relativistic quantum fields, except that the infinitely many degrees of freedom of a physical field operator like $\hat {\Psi} (x^\alpha)$ or  $\hat {A}_\mu (x^\alpha)$  and special relativity bring their own subtleties and complications. In this quantization program, firstly the canonical momentum density field and the energy momentum tensor field are obtained from the classical Lagrangian density of a field e.g. $\mathcal{L}_{D0}$ for the fermionic field $\Psi $. Then, these fields are elevated to the status of field operators by setting up appropriate canonical commutation brackets   between the field operators and their canonical momentum operators.

 Once the Hamiltonian density operator,  $\hat{\mathcal{H}}(\vec{r})$ (e.g. $\hat{\mathcal{H}}_{EM} = \frac{1}{8\pi}(\hat{E}^2 + \hat{B}^2) $,  in the case of EM fields),  is derived from the Lagrangian density using the canonical prescription, the time evolution of any physical state-vector, $\vert \psi >$,  can be determined from the  Schr$\ddot{\mbox{o}}$dinger picture,
\begin{equation}
i \hbar \frac {d \vert \psi>} {d t} = \hat{H} \vert \psi (t)> = \int{ \hat{\mathcal{H}}(\vec{r}) d^3r} \  \vert \psi (t)>
\end{equation}
where $\hat{H} \equiv  \int{ \hat{\mathcal{H}}(\vec{r}) d^3r} $ is the Hamiltonian operator.

(In quantum mechanics, the wavefunction $\psi(\vec{r},t)$ is simply the inner product
 $(|\vec{r}>, |\psi (t)>)$ = $\ <\vec{r}\ |\psi (t) >$,  with $|\vec{r}> $ being an eigenvector of the position operator $\hat{\vec{r}} $ corresponding to the eigenvalue $\vec{r}$.)
 
If the Hamiltonian $\hat{H}$ does not depend on time (as in the case of free fields) then for any initial state $\vert \psi (0)>$, it is straightforward to  show from eq.(16) that the state undergoes a unitary evolution generated by the Hamiltonian,
\begin{equation}
\vert \psi (t)> =  e^{- \frac {i}{\hbar} \  \hat{H}\  t} \  \vert \psi (0)> = e^{- \frac {i}{\hbar} \  \  t\ \int{ \hat{\mathcal{H}}(\vec{r}) d^3r}} \  \vert \psi (0)>
\end{equation}
In such a situation, one may adopt a Heisenberg picture description wherein the state-vectors remain the same and only the physical observables change with time. But when time-dependent interaction terms are present so that eq.(16) takes the form,
\begin{equation}
i \hbar \frac {d \vert \psi>} {d t} = (\hat{H} + \hat{H}_{int}(t))\vert \psi (t)> = \int{ (\hat{\mathcal{H}}(\vec{r}) + \hat{\mathcal{H}}_{int}(\vec{r}, t)) d^3r} \  \vert \psi (t)> \ \ \ ,
\end{equation}
it is convenient to use the interaction picture in which the state-vector $|\psi (t)>_I$  is define by,
\begin{equation}
|\psi (t)>_I \ \equiv \ e^{+ \frac {i}{\hbar} \  \hat{H}\  t} \  \vert \psi (t)> \ \ \ ,
\end{equation}
where $|\psi (t)>$ is the usual Schr$\ddot{\mbox{o}}$dinger picture state-vector.  From eq.(19), it is obvious that the state-vector $|\psi (t)>_I$ satisfies the following equation,
\begin{equation}
i \hbar \frac {d \vert \psi>_I} {d t}= \hat{H}_{int}(t)\vert \psi (t)>_I = \int{ \hat{\mathcal{H}}_{int}(\vec{r}, t) d^3r}\ \ \ \vert \psi (t)>_I
\end{equation}
In the case of QED, using eq.(14), one may express the interaction Hamiltonian density operator as,  $\hat{\mathcal{H}}_{int}(\vec{r}, t) = \frac{1} {c}\ \hat{j}^\mu \ \hat{ A}_\mu$.

But eqs.(20)  corresponds to a particular  inertial frame S for which a time $t$ has been established everywhere in S by synchronizing all the distributed clocks in the frame. If one wishes to express the evolution of a state-vector in a manifestly covariant manner, one needs to express the evolution with respect to space-like 3-dimensional hypersurfaces $\Sigma $. Tomonaga-Schwinger formalism had made use of functional derivatives with respect to hypersuface considering an infinitesimal deformation of the space-like hypersurface $\Sigma $ to obtain an evolution equation in the interaction picture of the form,
\begin{equation}
i \hbar \bigg ( |\psi (\Sigma + \delta \Sigma)>_I -  |\psi (\Sigma)>_I \bigg ) = \frac{1} {c} \bigg ( \int^{\Sigma + \delta \Sigma}_\Sigma {\hat{\mathcal{H}}_{int}\  d^4x}\bigg )\  |\psi (\Sigma)>_I
\end{equation}
which is manifestly covariant as can be deduced from the integral over the space-time proper volume. 

Schwinger had presented a sophisticated covariant quantization scheme for fields   in great detail [12]. The axiomatic style of the paper that systematically takes up various subtle issues in a complete and coherent manner is simply awesome. But because of the mathematical rigor, Schwinger's papers on QED appear formidable too.  

Schwinger  had adopted the interaction picture wherein the state-vector representing a system of photons and charged, spin 1/2 fermions evolve from one space-like hypersurface to another  only due to the interaction part of the Hamiltonian similar to eq.(21). Both Schwinger and Dyson had emphasized in their papers that all space-time dependent physical observables (i.e. hermitian operators)  must mutually commute and thereby be measurable on any space-like hypersurface. The modern name for this concept is the principle of micro-causality.  Micro-causality is a consequence of combining both quantum theory and special relativity, as argued below. 

It is well established from experiments that when an observable $\hat{O}$ is measured very accurately, the state-vector collapses immediately to one of former's eigenvectors. Operators that commute have simultaneous eigenvectors, and thus, a set of mutually commuting observables  can all be measured simultaneously and precisely.  This is the reason why, in an ideal and precise measurement, position and momentum cannot be measured simultaneously,  as  there is no common eigenvector of $\hat{x}$ and $\hat{p}$ since $[\hat{x}, \hat{p}]= i \hbar$. 

Now, on a space-like hypersurface $\Sigma$, no two events: $E_1$  at $(t_1, x_1, y_1, z_1)$ and $E_2$ at $(t_2, x_2, y_2, z_2)$,  can be causally connected  as $\ \ c^2 (t_1 -t_2)^2 - (x_1 -x_2)^2 - (y_1 -y_2)^2 - (z_1 -z_2)^2 < 0 \ \ $ on $\Sigma$. Even a ray of light would not be able to link $E_1$ and $E_2$. Hence, one ought to be able to measure any two physical quantities $\hat {O}_1$ at $E_1$ and $\hat {O}_2$ at $E_2$ accurately, since the measurements cannot influence each other  as they lie outside each other's light cone.

 Therefore, simultaneous eigenstates of $\hat {O}_1(E_1)$ and  $\hat {O}_2(E_2)$  must exist. This implies that  $[\hat {O}_1(E_1) , \hat {O}_2(E_2)]= 0$, which  precisely is the statement of micro-causality principle. In particular, even the commutator of a field operator at $E_1$ and its canonically conjugate momentum field operator at $E_2$ must vanish if $E_1$ and $E_2$ are space-like separated.

Although Dyson's starting point was Schwinger's approach, he simplified the mathematical apparatus employed by Schwinger  considerably without losing track of the physical processes like higher order radiative reactions and,  moreover, he went beyond Schwinger's formulation [6]. In order to solve eq.(21), Dyson introduced perturbation theory and employed a 1-parameter family of space-like hypersurfaces to obtain a unitary operator which in the asymptotic limit reduces to Heisenberg's  S-matrix. In a follow up paper, Dyson  obtained integral equations for Green's functions that ensue from  field equations [21].   Schwinger, employing alternate methods, arrived  independently at the same conclusion [22]. This set consisting of  an infinity of hierarchy equations go by the name of Dyson-Schwinger equations.   

While in Feynman's formulation of QED, as discussed earlier, every physical process is assigned a probability amplitude in accordance with the principles of quantum theory applied to the Dirac equation (after switching on the EM fields i.e. eq.(9)). It is not surprising that the space-time approach  appeared  very  dissimilar to the  Tomonaga-Schwinger formalism, whose focus among other things was to obtain covariant unitary evolution operators  based on dynamical fields,  until Dyson arrived at the scene. He provided a very systematic formulation of Feynman's approach in terms of S-matrix theory [6, 21]. 

 Not only did he prove the equivalence of Feynman and Tomonaga-Schwinger formulations, he also showed that the renormalization technique gives finite result at all orders of the perturbation theory. For QED, Dyson was the first person to lay out an exhaustive treatment of renormalization [21].  These were the  crucial steps that led the 1965 Nobel prize in physics be awarded to  Tomonaga, Schwinger and Feynman. 
 
\section{Ferromagnets, Stability of Matter, Gravity  and  Dyson Bound}

Around 1950s, Dyson was drawn to the subject of  ferromagnets and spin waves. Ferromagnets are basically  substances in which  magnetic dipole moments at the lattice sites get aligned in the direction of an applied magnetic field. These magnetic dipole moments associated with atoms (like  iron, nickel  or cobalt) arise due  to their uncompensated electronic spins. In a seminal study, Felix Bloch  had introduced the notion of spin waves in 1930, and had argued that  the  spin wave degrees of freedom are important for  ferromagnetism  at  temperatures far below the Curie temperature [23]. 

As the name suggests, spin waves (the corresponding quanta being called the magnons) are essentially   undulating changes  in the  orientation of magnetic dipole moments  propagating along the lattice sites of a ferromagnetic crystal. Following Bloch's insight,  calculations were done subsequently by several authors pertaining to the spontaneous magnetization at low temperatures,  beyond the leading order term with a temperature dependence of $ T^{3/2}$ that Bloch [23] had  obtained. But these studies   led to conflicting reports on the thermodynamics aspects of ferromagnets, until Dyson's  two back-to-back papers appeared on this subject in 1956. In the first paper, Dyson addressed the problems posed by the non-orthogonal spin-wave states and provided a suitable formalism  to calculate thermodynamic quantities, in terms of these states in the context of a ferromagnet [24]. 

He also studied the collision of spin waves and, in particular, gave an exact expression for the free energy that incorporated the effects of spin wave interactions, for he  had also calculated the scattering cross-section for two interacting spin waves at low temperatures. In the second paper, Dyson demonstrated the flaws in the earlier papers and calculated the free energy of  a Heisenberg model for an ideal ferromagnet,  using the techniques developed in his first paper [24] and showed that  the effects of interacting spin waves in the spontaneous magnetization begins at order with temperature dependence of  $T^4$ [25].

In 1960s, inspired by an address of P. Ehrenfest to W. Pauli that, `...(why) matter should occupy so large a volume... But why are the atoms themselves so big?', Dyson turned his attention to the study of stability of macroscopic systems.  Dyson \& Lenard [26], Dyson [27] and Lenard \& Dyson [28] were the first to prove rigorously some  powerful theorems concerning the importance of Pauli's exclusion principle in the stability of bulk matter made up of non-relativistic fermions interacting with each other through  electrostatic  forces. 

For a  quantum system made up of N negatively charged fermions and arbitrary number of positively charged particles, they proved that  there exists a lower bound $E_0$ to the minimum energy $E_{min}$ so that,
\begin{equation}
E_{min} > E_0 \propto   - N \bigg (\frac{m_e e^4}{2 \hbar^2} \bigg )
\end{equation}

On the other hand, by considering a non-relativistic quantum system consisting of equal number $N$ of positively and negatively  charged bosons all with same  magnitude of charge $|e|$ as well as identical electronic mass $m_e$, Dyson proved that if the interactions are purely Coulombic then the ground state energy of the Hamiltonian has an upper bound [27],
\begin{equation}
E_N < - \frac{1} {1944 \pi^4}  \bigg (\frac{m_e e^4}{2 \hbar^2} \bigg ) N^{7/5}\ \ \ ,
\end{equation}
indicating that the system is likely  to be unstable. Hence, the results given by eq.(22) and eq.(23) underscore the significant role that Pauli's exclusion  principle plays in ensuring the stability of macroscopic systems and, therefore, of the entire physical world.

According to eq.(23), for  bulk matter containing an Avogadro number of such bosons, the energy released while going to the ground state would be at least $\sim   10^{23}$ erg ! Interestingly enough, if one considers an astronomically large number of identical, ultra-light  bosons interacting only via  Newtonian gravity,  one can show that such a system is likely to collapse gravitationally into a black hole [29, 30]. 

Just after publishing his papers on the stability of bulk substances, Dyson shifted his gaze at the topic of  gravitational waves (GWs). In any relativistic theory of universal gravitation, if energy and momentum (both being the source of gravity) of or within  a body change in an asymmetric manner, the resulting changes in the gravitational effects would necessarily have to propagate in the form of a wave with speed $\leq c$. Simply put, this is how GWs are generated. In Einstein's general relativity (GR), GWs are produced whenever mass quadrupole moment tensor associated with a source changes with time (see e.g. [30]).  

Spurred by J. Weber's theoretical work on GWs as well as his stupendous efforts in building a  very sensitive resonant bar detector to observe GWs, Dyson calculated the effect of GWs having frequencies in $\sim $ 1 Hz band on bodies like Earth assuming them to be elastic solids [31].  Dyson incorporated compression, shear-wave  as well as rotation in his analysis and showed that the response to GWs depended on the non-uniformities in the shear-wave modulus. He found that GWs are absorbed by an elastic  object when the shear-wave modulus is inhomogeneous. This was the first time that excitations of the normal modes of oscillations of an astrophysical body, like Earth, due to incident GWs was being addressed.

However, in the field of GWs studies,  Dyson's name would  forever be associated with a conjecture on an  upper bound on GW luminosity ([32];  but also see  the references provided in [30], on this topic). The quantity $\frac{c^5} {G} = 3.6 \times 10^{59} \ \mbox{erg s}^{-1}\ \ \equiv L_{Dyson}$, which has the dimension of luminosity, is called the Dyson bound on the GW luminosity. It is generally surmised that no GW source can radiate energy at a rate exceeding $L_{Dyson}$.

It is somewhat surprising to note that $L_{Dyson}$  emerges also when one considers the well known Planck energy, $E_{Pl} \equiv m_{Pl} c^2 \equiv \sqrt{c^5 \hbar / G}$  and the Planck time, $t_{Pl} \equiv \sqrt{\hbar G/ c^5}$ to obtain a Planck luminosity [30],
\begin{equation}
\frac{E_{Pl}} {t_{Pl}} = \frac{c^5} {G} \ \  \ ,
\end{equation} 
since the Planck constant gets cancelled. Now, quantum fluctuations could have generated many forms of radiation at the time of the initial big bang with a characteristic luminosity $\frac{E_{Pl}} {t_{Pl}}  = L_{Dyson} $, but only the classical gravitational signature from GR survives in this expression.  There is perhaps an underlying subtlety  in the concept of Dyson bound that is yet to be fathomed.

\section{Final Leaps of Eternal Intelligence, Trees and Spheres in order to Disturb the Cosmos}

After going through  Jamal Islam's `Possible ultimate fate of the universe' [33], Dyson was induced  into   contemplating on  the very distant  future of the cosmos that follows from the big bang model.  Charged with an insight that intelligent species could influence the fate of the evolving universe, Dyson published an exceedingly thought provoking  review paper - `Time without end: Physics and biology in an open universe' [34]. 

This article is  essentially based on the   "James Arthur
Lectures on Time and its Mysteries"  that he had delivered at New York University, in 1978. It begins with reviewing some of the seminal works of Marteen Rees and Steven Weinberg but quickly goes into Dyson's original ideas pertaining to the remote cosmic future.

As the second law of thermodynamics (i.e. monotonic increase in entropy of an isolated system) is  invincible, Dyson asked whether it is possible to maintain life and intelligence with the stellar sources of energy (necessary for food production and  metabolic activities)  fading away eventually in an ever expanding universe? Dyson explored the possible manoeuvres that could be undertaken by a super advanced civilization to survive as well as to continue thinking, albeit intermittently as the `intelligent being or automaton' would need to go into long hibernation to ration dying energy resources. This idea is often referred to as Dyson's Eternal Intelligence.

He discussed such futuristic concepts even in his autobiographical book `Disturbing the Universe' [35]. Dyson's prodigious output is unique and remarkable because not only he proved rigorous theorems in diverse fields of  physics based on hard and cold mathematical calculations, he also made imaginative but quantitative speculations concerning remote future. For instance, taking into account the presence of water in comets, he speculated on the possibility of growing genetically engineered plants on approaching comets (the so called Dyson Tree).

 Similarly, noting that most of the energy radiated away almost isotropically by stars go wasted, advanced civilization on habitable planets like Earth could increase the efficiency of harnessing the stellar energy by surrounding their host stars in a near isotropic manner by orbiting asteroid like objects mounted with devices (e.g. `stellar panels') to absorb the stellar radiation for useful purposes. 
 
 One of  the consequences of having such a `Dyson Sphere' (DS) installed by an extraterrestrial civilization around its star would be that the constituent objects of the DS would re-radiate in the infrared part of the visible spectrum, and therefore such IR emission would be a likely signature of the presence of extraterrestrial intelligence in exoplanets. Dyson's thoughts must be echoing around the frenetic activities currently taking place pertaining to the spate  of discoveries of exoplanets.
 
\section*{Acknowledgements}
It is a pleasure to thank Professor B. Sury for his constant encouragement to write on Late Professor Freeman Dyson's outstanding contributions in the field of physical sciences.

\section*{References}

\begin{enumerate}
\item [[1]] H. A. Bethe, Phys. Rev., Vol. 72, p. 339, 1947.

\item [[2]] F. J. Dyson, Physics Today, Vol. 58, p. 48,  2005.

\item [[3]] A.N. Mitra, Resonance, Vol. 10, p. 33, 2005.

\item [[4]] F. J. Dyson,   Phys. Rev., Vol.73, p. 617, 1948.

\item [[5]] J. Schwinger, Phys. Rev., Vol.73, p. 415, 1948.

\item [[6]] F. J. Dyson,   Phys. Rev., Vol.75, p. 486, 1949.

\item [[7]] F. J. Dyson, `Maker of Patterns: An Autobiography Through Letters', Liveright Publishing Corporation, 2018.

\item [[8]] Sin-itiro Tomonaga, Prog. Theor. Phys., Vol. 1, p. 27, 1946.

\item [[9]] Z. Koba, T. Tati and S. Tomonaga, Prog. Theor. Phys., Vol. 2, p. 101, 1947.

\item [[10]] S. Kanesawa and S. Tomonaga, Prog. Theor.
Phys., Vol. 3, p. 101, 1948.

\item [[11]] S. Tomonaga and J. R. Oppenheimer, Phys. Rev.,  Vol. 74, p. 224, 1948.


\item [[12]] J. Schwinger,  Phys. Rev., Vol. 74, p. 1439, 1948.

\item [[13]] J. Schwinger, Phys. Rev., Vol. 75, p.651, 1949.

\item [[14]] J. Schwinger, Phys. Rev. Vol. 76, p.790, 1949.


\item [[15]] R. P. Feynman, Phys. Rev. Vol.74, 1430, 1948.

\item [[16]] R. P. Feynman, Phys. Rev. Vol. 76, 749, 1949.

\item [[17]] R. P. Feynman, Phys. Rev. Vol. 76, 769, 1949.

\item [[18]] V. B. Berestetskii, E. M. Lifshitz \& L. P. Pitaevskii, `Quantum Electrodynamics', Pergamon Press, 1982.

\item [[19]] P. Das Gupta, in `Lectures on Quantum Mechanics: Fundamentals and Applications', eds. Anirban Pathak and Ajoy Ghatak, Viva Books Pvt. Ltd., 2019.

\item [[20]] E. C. C. St$\ddot{\mbox{u}}$ckelberg, Helv. Phys. Acta., Vol. 15, p.23, 1942.

\item [[21]] F. J. Dyson,   Phys. Rev., Vol.75, p. 1736, 1949.

\item [[22]] J. Schwinger, Proc. Nat. Acad. Sc. Vol. 37, 452, 1951. 

\item [[23]] F. Bloch,  Z. Phys., Vol. 61, p. 206, 1930.
\item [[24]] F. J. Dyson,  Phys. Rev., Vol. 102, p.1217, 1956.

\item [[25]] F. J. Dyson, Phys. Rev., Vol. 102, 1230, 1956.

\item [[26]] F. J. Dyson  \& A. Lenard, J. Math. Phys., Vol. 8, p.423, 1967.

\item [[27]] F. J. Dyson, J. Math. Phys., Vol. 8,  p. 1538, 1967.

\item [[28]] A. Lenard  \& F. J. Dyson, J. Math. Phys., Vol. 9, p. 698, 1968.

\item [[29]] P. Das Gupta \& Eklavya Thareja, Class. Quan. Grav., Vol. 34, 2017.

\item [[30]] P. Das Gupta \& Fazlu Rahman, in `The Physical Universe-2018’, eds.  S. Wagh, S. Maharaj \& G. Chon (CIRI, Nagpur, 2018);  arXiv:1801.02559v2 [gr-qc]

\item [[31]] F. J. Dyson, Astrophys. J., Vol. 156, p.529, 1969.

\item [[32]] F. Dyson, in Interstellar Communication, ed. A.G. Cameron, (New
York: Benjamin, 1963), chap 12.

\item [[33]]  J. N. Islam, Quart. Jour. Roy. Astron. Soc.,  Vol. 18, p.3, 1977.

\item [[34]] F.  J. Dyson, Rev. Mod. Phys., Vol.  51, p. 447, 1979.

\item [[35]]  F. Dyson, `Disturbing the Universe',  Harper and Row, New York, 1979.
\end{enumerate}

\end{document}